\documentclass[conference]{IEEEtran}

\usepackage{cite}      

\usepackage{graphicx}  

\usepackage{psfrag}    

\usepackage{subfigure} 

\usepackage{url}       

\usepackage{amsmath}   
\begin{document}

\title{Gamma Ray Astronomy \\ and the Origin of Galactic Cosmic Rays}

\author{\authorblockN{Stefano Gabici}
\authorblockA{School of Cosmic Physics, Dublin Institute for Advanced Studies\\
31 Fitzwilliam Place, Dublin 2, Ireland\\
Email: sgabici@cp.dias.ie}
}


%


\maketitle

\begin{abstract}
Diffusive shock acceleration operating at expanding supernova remnant shells is by far the most popular model for the origin of galactic cosmic rays. Despite the general consensus received by this model, an unambiguous and conclusive proof of the supernova remnant hypothesis is still missing.
In this context, the recent developments in gamma ray astronomy provide us with precious insights into the problem of the origin of galactic cosmic rays, since production of gamma rays is expected both during the acceleration of cosmic rays at supernova remnant shocks and during their subsequent propagation in the interstellar medium.  
In particular, the recent detection of a number of supernova remnants at TeV energies nicely fits with the model, but it still does not constitute a conclusive proof of it, mainly due to the difficulty of disentangling the hadronic and leptonic contributions to the observed gamma ray emission.
In this paper, the most relevant cosmic--ray--related results of gamma ray astronomy are briefly summarized, and the foreseeable contribution of future gamma ray observations to the final solution of the problem of cosmic ray origin is discussed.
\end{abstract}


%
\IEEEpeerreviewmaketitle

\section{INTRODUCTION}

Since their discovery in 1912, cosmic ray (CR) nuclei and electrons have been studied by means of constantly improving direct and indirect detection techniques \cite{gaisser}.
Despite exciting experimental results and extensive theoretical efforts over
the past decades, the origin of these particles is still debated. The problem is that, unlike photons, CRs
are deflected and isotropized by the galactic magnetic field and thus their arrival direction does not
point back to the actual position of their accelerators. Revealing the mystery of their origin is of
fundamental importance, since CRs can provide unique information about the physical conditions of
the extreme astrophysical objects in which they are likely to be accelerated. Moreover, the energy
density of CRs, largely dominated by the hadronic component, is comparable with the pressure of the galactic magnetic field as well as to that of the
interstellar medium. This implies that CRs have an important role in the dynamical balance of our
Galaxy \cite{parker} and influence interstellar chemistry through the heating and ionization of interstellar gas \cite{palla}.

In 1934, Baade and Zwicky first proposed that supernovae are the sources of galactic CRs \cite{zwicky}. To
support their idea, they used a simple argument: the observed CR population can be maintained at
the present level if a small fraction (a few percent) of the galactic supernovae kinetic energy is
somehow converted into CRs. This argument is strengthened by the fact that it is commonly believed
that CRs can be efficiently accelerated via Fermi mechanism at shock waves that form during the
expansion of supernova remnants (SNRs) in the interstellar medium (see \cite{drury,blandford,jones,malkov} for reviews).
The acceleration of CRs in SNRs must be accompanied by a copious gamma ray
emission due to the decay of neutral pions produced in interactions between relativistic protons and
protons in the interstellar medium. Because the energy transferred to accelerated particles is tightly
constrained by the observed total CR power of the Galaxy, it is possible to obtain an almost model independent
prediction of the gamma ray luminosity of SNRs \cite{dav}. 

To date, several SNRs have been detected at TeV energies by the major currently operating Cherenkov telescopes
(e.g. \cite{RXJnature}, \cite{rev:felix, rev:dieter, rev:jim, rev:persic, grenierICRC} for reviews), whose flux level matches very well the above mentioned
predictions. Though these results undoubtedly constitute one of the most important
advancements in the field, they still do not provide us with a definite and direct evidence of proton
acceleration at SNRs. In fact, competing leptonic processes can also explain the observed TeV gamma
ray emission, provided that the magnetic field does not exceed $\approx 10 ~ \mu$G, and thus accurate modelling is needed in order to disentangle the different contributions.
Evidence for strong $\approx 100 ~ \mu$G magnetic field, and thus indirect support  to the hadronic scenario for the gamma ray emission, comes from the observation of thin  X-ray synchrotron filaments surrounding some SNRs \cite{yasfirst,bamba,vink} and from the observed rapid variability of the synchrotron X-rays from the SNR RXJ1713.7-3946 \cite{yasunobu}.
Also neutrinos are produced during the hadronic interactions responsible for the generation of
gamma rays. Their detection, though challenging even for the next generation of telescopes, would
constitute an unambiguous proof for proton acceleration in SNRs \cite{dav,vissani,kappes,meapj,halzen}.

Another difficulty of the supernova hypothesis for the origin of galactic CRs is related to the very low anisotropy observed in the arrival direction of CRs up to energies of $10^{15}$eV and above \cite{CRbook}.
This is in conflict with the extremely short CR confinement time in the Galaxy (which would imply anisotropy) required to steepen the CR spectrum from the hard injection spectrum  ($\approx E^{-2}$) predicted by diffusive shock acceleration to the steep spectrum ($\approx E^{-2.7}$) observed locally (for a discussion see e.g. \cite{hillas}). 
To solve this problem, not only the details of the acceleration mechanism operating at SNR shocks needs to be fully understood, but also the way in which CRs are released into the interstellar medium during the SNR evolution and the way in which they subsequently propagate in the Galaxy.

It is believed that CRs accelerated at SNR shocks are gradually released in the interstellar medium as the SNR expand and the shock slows down \cite{ptuskin}. The bulk of these CRs remains diffusively confined within the Galaxy for about ten millions years before escaping.
During this time CRs undergo inelastic interactions in the interstellar gas and produce neutral pions which decay into gamma rays (see e.g. \cite{CRbook}). The observed galactic diffuse GeV emission which exhibit a good spatial correlation with the gaseous disk of the Galaxy is the result of these CR interactions \cite{EGRETdiffuse}.
On much smaller scales, a correlation between TeV diffuse gamma ray emission and gas density has also been observed from regions of the galactic disk characterized by the presence of massive molecular cloud complexes: the galactic centre ridge \cite{hessridge} and the Cygnus region \cite{milagrocyg}. Such correlations are generally considered as hints for a hadronic origin of the gamma ray emission, since the presence of massive molecular clouds provide a dense target for CR hadronic interactions and thus enhances the expected gamma ray emission.
A leptonic origin, though not ruled out (especially for the Cygnus region \cite{milagrocyg}), seems disfavored since it would require fine tuning to explain the observed correlation. 
For the same reason, the detection of TeV gamma rays from a few SNRs spatially associated with dense molecular clouds \cite{IC443,W28,HESSJ1745,CTB} supports the idea that such TeV emission has a hadronic origin and that SNRs might indeed be the sources of galactic CRs.

Even in the absence of a nearby CR accelerator, molecular clouds embedded in the "sea" of galactic CRs are expected to emit gamma rays. If CRs can freely penetrate the cloud, the resulting gamma ray spectrum is expected to mimic the underlying CR spectrum. For this reason, molecular clouds can be used as "CR barometers" to probe variations of the CR spectrum and flux throughout the Galaxy \cite{issa, felixclouds}. Such variations, if detected, will have to be accounted for by models describing the CR injection and propagation in the Galaxy (see \cite{GALPROPreview} for a review of propagation models).

Finally, independently on their nature, the accelerators of the galactic CRs are expected to emit gamma rays at some level and thus surveys of the galactic plane in gamma rays might reveal their presence.
Interestingly, more than a half of the low galactic latitude sources detected by EGRET at GeV energies \cite{3rdEGRET,olafunid} and many of the TeV sources discovered by H.E.S.S. \cite{HESSscan,HESSscan2} and by MILAGRO \cite{MILAGROsources} along the galactic plane still lack of any clear identification with objects at other wavelengths.
Revealing the mystery of the nature of the unidentified gamma ray sources might shed light on the problem of the origin of galactic CRs.

To conclude, despite the fact that that there is now encouraging convergence between the model and the observations, we are still waiting for the conclusive proof that galactic CRs are accelerated at SNRs. The difficulty in distinguishing between hadronic and leptonic contribution to the gamma ray emissivity of SNRs and the observed isotropy of the diffuse flux of CRs up to high energies constitute the two main problems to be solved.
Forthcoming gamma ray observations in both the GeV and TeV domain will constitute a fundamental step towards the solution of the problem of the origin of CRs.
In particular, deep ($\sim~5$~yr) FERMI/GLAST observations of SNRs will constrain emission models and possibly break the degeneracy of leptonic or hadronic origin of the gamma ray emission \cite{funkSNRs}. Moreover, a better determination of the spectrum and spatial distribution of the diffuse galactic GeV emission will improve our knowledge of the propagation of CRs and their distribution in the Galaxy. Similarly, TeV instruments of the next generation will be required to disentangle the emission form leptonic versus hadronic CRs by means of improved spectral measurements and morphological studies.
An extension of the observed energy range up to several hundred  TeVs is of crucial importance, since at those energies the contribution to the observed gamma ray emission from CR electrons is expected to be suppressed due to Klein-Nishina effects. 
Therefore, unlike other energy intervals, the interpretation of gamma ray observations at these energies is free of confusion and reduces to the only possible mechanism: decay of $\pi^0$ produced in hadronic interactions of PeV CRs \cite{meapj}.

The paper is structured as follows. In Sec.~2 the available gamma ray observations of SNRs and their theoretical interpretations are reviewed, with particular attention to the question on the leptonic or hadronic nature of the emission.
In Sec.~3 the importance of studying molecular clouds in gamma rays is highlighted, while Sec.~4 deals with the diffuse gamma ray emission observed along the galactic plane and the related issue of the spatial distribution of CRs in the Galaxy.
In Sec.~5 the capability of SNRs of accelerating CRs up to PeV energies is discussed, and a possible way to test this hypothesis is proposed.
Conclusions and future perspectives are provided in Sec.~6.

\section{GAMMA RAYS FROM SUPERNOVA REMNANTS: \\ HADRONIC OR LEPTONIC?} 

\begin{figure*}[ht]
   \centering
   \includegraphics[width=.45\textwidth]{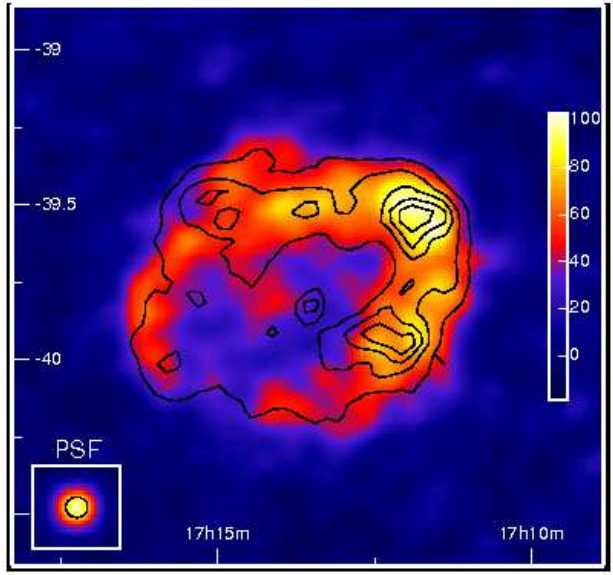}
   \hspace{0.5cm}
   \includegraphics[width=.45\textwidth]{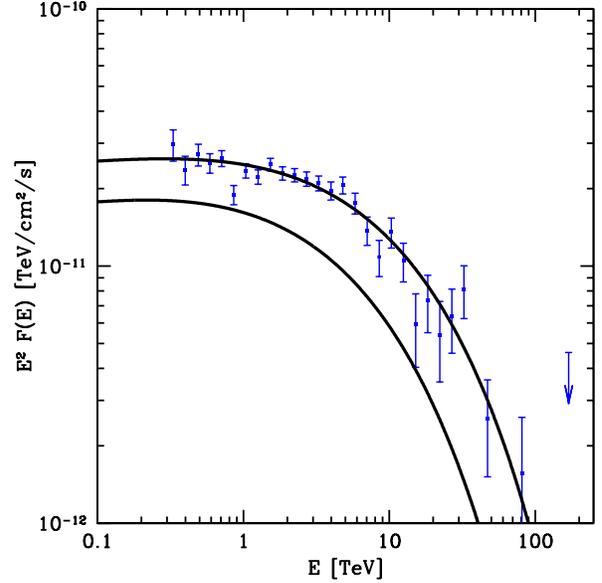}
   \caption{{\bf Left panel:} RX J1713.7-3946 as seen by H.E.S.S. (colors) and by ASCA in the $1 \div 3$
      keV energy band (contours). Figure from Ref.~\cite{RXJ100TeV}. A shell-like morphology is evident at both TeV and X-ray energies. {\bf Right panel:} spectrum of RX J1713.7-3945
      as measured by H.E.S.S. above 300 GeV. The upper solid line is a fit to the data by assuming that
      gamma rays have a hadronic origin and that the underlying CR proton spectrum is a power law in energy
      with slope 2 and an exponential cutoff at 150 TeV. The lower solid line represent the expected 
      neutrino flux from the same proton population.}
\end{figure*}

On average, every $\approx 30$ yr a supernova explodes in our Galaxy, releasing $\sim 10^{51}$ erg in form of mechanical energy \cite{SN}. If supernovae are the sources of galactic CRs, they have to convert $\approx 10 \%$ of this energy into relativistic protons in order to provide the total power of $\approx 10^{41}$ erg/s needed to maintain the galactic CR population at the observed level \cite{gaisser, hillas}.
Such conversion is likely to happen at the SNR shocks, where particles are believed to be accelerated up to ultrarelativistic energies via diffusive shock acceleration \cite{drury,blandford,jones,malkov}.
Once the total energy in accelerated protons per SNR is determined, it is possible to estimate the expected gamma ray luminosity due to hadronic interactions between relativistic protons and protons in the intertsellar medium swept up by the SNR shock.
If we assume a hard CR proton spectrum with differential energy distribution $N_{CR} \propto E^{-2}$, as suggested by shock acceleration theory, the expected gamma ray emission from a single SNR $J_{\gamma}(>E_{\gamma})$ will roughly follow the same energy distribution and will depend only on the total energy in accelerated protons $W_{CR}$, on the density of the ambient gas $n_{gas}$ and on the distance of the SNR $d$ as \cite{dav}:
\begin{equation}
J_{\gamma}(>E_{\gamma}) = 10^{-11} A \left( \frac{E_{\gamma}}{{\rm TeV}} \right)^{-1} {\rm cm}^{-2} {\rm s}^{-1} ,
\end{equation} 
$$
A = \left( \frac{W_{CR}}{10^{50} {\rm erg}} \right) \left( \frac{n_{gas}}{{\rm cm}^{-3}} \right) \left( \frac{d}{{\rm kpc}} \right)^{-2} .
$$
These expected fluxes are in general agreement with the fluxes of the SNRs observed at TeV energies.

To date, three young SNRs with apparent shell-type morphology have been detected by H.E.S.S. in TeV gamma rays: RX J1713.7-3946 (see Fig.~1, left panel) \cite{RXJnature}, RX J0852.0-4622 (Vela Junior) \cite{veladisc} and RCW 86 \cite{RCW}.
These observations demonstrate the capability of stereoscopic systems of performing detailed morphological studies of extended object at TeV energies with an angular resolution of $\sim 0.1^{\circ}$. This good angular resolution can be exploited to search for similarities in the observed morphologies at different wavelengths.
Notably, for both RX J1713.7-3946 and Vela Junior, X-ray observations revealed the presence of non-thermal synchrotron emission exhibiting a striking morphological similarity with the TeV gamma ray image \cite{RXJmorph,velamorph}.
Such a correlation is naturally expected in leptonic models, where both X-rays and gamma rays are emitted by the same population of electrons via synchrotron and inverse Compton scattering respectively.
On the other hand, the correlation can be accommodated also within hadronic models if most of the emission through both  $\pi^0$-decay gamma rays and synchrotron X-rays comes from regions characterized by high magnetic field and gas density \cite{rev:felix,BVlast}.
A scenario in which the X-ray emission is the result of synchrotron emission from secondary electrons from decay of charged pions would explain the correlation but cannot be invoked in this case, since the observed X-ray flux exceeds the flux of gamma rays, while one would expect the opposite to happen if the electrons were secondary products of proton interactions.

Another remarkable fact is that the gamma ray spectrum of RX J1713.7-3946 has been measured up to a photon energy exceeding 30 TeV (see Fig.~1, right panel), providing evidence for the existence of primary radiating particles (protons or electrons) with spectra extending above such energy \cite{RXJ100TeV}.
This demonstrate the ability of SNRs of accelerating particles up to at least $\approx 100$~TeV. Such an energy approaches the one of the CR {\it knee} ($\sim 4$ PeV), where the observed spectrum of CRs significantly steepens (see e.g. \cite{CRbook}). Due to the absence of any distinct feature in the CR spectrum below the knee, it is generally believed that the main contribution to the spectrum of galactic CRs comes from a single source population and that the representatives of such population must be able to accelerate particles up to {\it at least} a few PeV. 
Thus, any evidence that SNRs can act as {\it CR PeVatrons} will strengthen the hypothesis that they indeed are the sources of galactic CRs \cite{meapj}.
Future observations in the multi-TeV energy range \cite{tenten} will possibly solve this issue (see Sec.~5 for a discussion).

Apart from the three objects mentioned above, the list of SNRs detected at TeV energies includes also Cas A \cite{CasAHEGRA,CasAMAGIC,VeritasSNRs}, IC443 \cite{IC443,VeritasSNRs} and SN 1006 \cite{SN1006}.
Good candidates for TeV SNRs are HESS J1713-381 (likely to be associated to the SNR CTB 37B) \cite{CTB37B}, HESS1714-385 (CTB A) \cite{CTB} and HESS J1745-303 (a possible SNR/molecular cloud association) \cite{HESSJ1745}.

The question of the hadronic or leptonic nature of the TeV emission from SNRs constitutes one of the most discussed issues in gamma ray astronomy.
If, from the one side, hadronic TeV gamma rays are expected if SNRs are the sources of galactic CRs, it is also true that most of the SNRs are sources of non-thermal X-rays, commonly interpreted as synchrotron radiation from multi-TeV electrons (see e.g. \cite{reynolds,vladfelix,reynoldsrev}).
Such electrons can also emit TeV gamma rays via inverse Compton scattering off photons of the cosmic microwave background, thus providing a competing emission process responsible for the TeV radiation (see Fig.~2 for a leptonic model for RX J1713.7-3946).
If the magnetic field at the SNR is stronger than $\approx 10 ~ \mu$G, then the observed synchrotron X-rays can be explained by a relatively meagre number of electrons, which would produce unappreciable TeV inverse Compton emission. Conversely, if the value of the magnetic field is $\approx 10 ~ \mu$G the electrons needed to explain the X-ray emission will also suffice to explain the whole observed TeV emission, thus implying inefficient acceleration of CR protons. 
Thus, the value of the magnetic field at the shock is a crucial parameter of the problem, and its determination would allow us to unveil the nature of the gamma ray emission.

\begin{figure*}[ht]
   \centering
   \includegraphics[width=.7\textwidth]{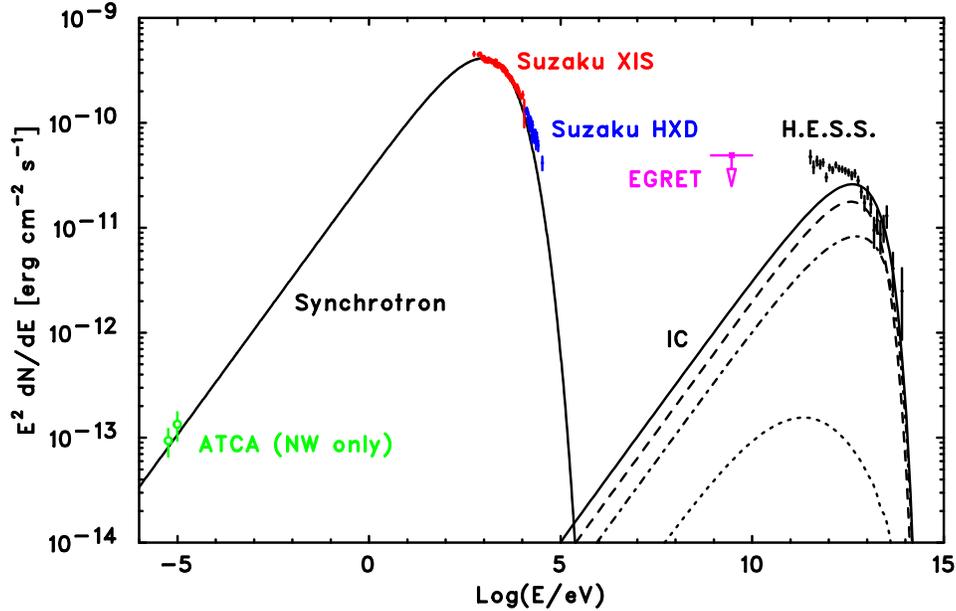}
   \caption{Observed spectrum of RX J1713.7-3946 (data points) with a leptonic model (solid lines). Synchrotron and Inverse Compton (IC) emission are shown for an injection spectrum of electrons which is a power law in energy with slope 2. An excellent fit to the X-ray data is obtained by assuming that the shape of the cutoff in the electron spectrum is the one expected from diffusive shock acceleration in presence of synchrotron losses \cite{vladfelix}. Magnetic field is 14 $\mu$G. The inverse Compton flux has been calculated by using the interstellar radiation model from Ref.~\cite{porter}. The model fails to fit the low energy part of the measured gamma ray spectrum. Figure from Ref.~\cite{tanaka}.}
\end{figure*}

The observation of thin (down to a few arcseconds scale) synchrotron X-ray filaments surrounding a number of SNRs has been interpreted as an evidence for the presence of a strong ($\approx 100 ~ \mu$G or more) magnetic field at the shock (see e.g. \cite{bamba,yasfirst,vink,gamiltycho,bambahist,warren,lazendic}).
According to this interpretation, the formation of such filaments is due to the fact that, in such a strong magnetic field, the synchrotron cooling time of X-ray emitting electrons is very short. Thus electrons radiate X-ray synchrotron photons before being significantly advected or diffuse away downstream of the shock \cite{volkfilaments,donfilaments,parizotfilaments}.
The value of the magnetic field is estimated by comparing the observed width of the filament with the expected one, which is roughly $\approx \sqrt{k_d \tau_d}$, where $k_d$ and $\tau_d$ are the diffusion coefficient and synchrotron loss time of the emitting electrons, both quantities depending on the value of the magnetic field \cite{volk1006}.
The derived value of the magnetic field relies on the (very plausible) assumption that diffusion is proceeding at the Bohm rate, even though attempts to use X-ray data to derive the diffusion coefficient have been made \cite{parizotfilaments,bohmnature}.
However, such estimates \cite{parizotfilaments,bohmnature} are seriously affected by the uncertainty in the determination of supernova parameters such as the shock speed and, most of all, the cutoff energy in the synchrotron spectra.

The presence of a high magnetic field at the shock indirectly supports the hadronic origin of the TeV emission for two main reasons.
First, according to theoretical studies, magnetic field amplification due to non-resonant CR driven instability is expected to take place at shocks which are accelerating efficiently CR protons \cite{lucek,bell}.
The predicted value of the amplified field, higher than earlier estimates based on CR generation of resonant Alfven modes \cite{lagage}, is compatible with the values inferred from X-ray observations and it is also the value needed to allow acceleration to proceed up to the energy of the CR knee or even above \cite{bell}. 
Second, as said above, if the magnetic field is stronger than $\sim 10 ~ \mu$G the inverse Compton scattering contribution to the TeV emission is negligible, leaving the hadronic channel as the only viable mechanism for gamma ray production.

A high value of the magnetic field is not the only possible interpretation for the observed narrow X-ray synchrotron filaments. Filaments can also be formed due to damping of magnetic turbulence downstream of the shock \cite{pohl}. 
In this scenario, though magnetic field amplification can still operate at the shock , the width of filaments can not be used to estimate the value of the magnetic field, instead it simply reflects the spatial structure of the magnetic field.
An important difference between the magnetically limited filaments and the energy loss limited filaments is that in the former case filamentary structures, though broader and of less amplitude, should be observed also at lower (e.g. radio) frequencies \cite{pohl}, while in the latter case no filaments are expected at radio frequencies since low energy electrons are unaffected by synchrotron losses \cite{donfilaments,gamilfilaments}.
In the case of the SNR SN~1006 the profile of the radio emission is quite broad, with only shallow bumps, when compared to the narrow peak in the X-ray emission \cite{1006radio}. 
However, an accurate comparison between observation and theory is needed in order to draw firm conclusions.
Similarly, a comparison between the X-ray and radio morphology of the Tycho SNR has been attempted, but results still remain inconclusive \cite{gamiltycho}.

Other evidences for the presence of a strong magnetic field at SNR shocks comes from the recently discovered fast variability of synchrotron X-rays hot spots in the shell of RX J1713.7-3946 \cite{yasunobu} and Cas A \cite{yasunobu2}.
An extremely strong magnetic field of $\approx 1$ mG has been inferred by comparing the decay time of the X-ray flux, of the order of  a year, with the synchrotron cooling time.
Though the estimate of the magnetic fields refers only to the sub-parsec scale hot spot regions which are seen to vary in X-rays, it suggests that significant amplification of the field can possibly happen throughout the whole SNR shell.

Finally, another way to discriminate between hadronic or leptonic origin of the TeV emission consists in comparing model predictions with the observed X-ray and TeV gamma ray spectra. 
While hadronic models based on predictions of non-linear shock acceleration theory generally fit quite satisfactorily both X-ray and gamma ray data (for RX J1713.7-3946 see e.g. \cite{volkRXJ,morlino,BVlast} and right panel of Fig.~1 for a fit of H.E.S.S. data based on hadronic interactions), leptonic models seem to provide worse fits (see Fig.~2).
High quality SUZAKU data have been obtained over two decades in energy ($\approx 0.5 \div 50$ keV) for the SNR RX~J1713.7-3946 \cite{tanaka}. 
Such data allow us to constrain the spectrum of the synchrotron emitting electrons with unprecedented accuracy and over a wide energy range.
Remarkably, the resulting electron spectrum is very close to that expected fom electrons accelerated via diffusive shock acceleration in presence of synchrotron losses (see Fig.~2) \cite{vladfelix,tanaka}.
The electron spectrum derived in this way can be used to calculate the TeV spectrum due to inverse Compton emission off photons of the cosmic microwave background and of the galactic optical and infrared background.
It is clear from Fig.~2 that such an inverse Compton spectrum fails to fit the lowest energy part of the measured TeV spectrum, even assuming an optimistic model for the interstellar radiation field \cite{porter}.
The fit can be improved only at the expense of enhancing the infrared background of a factor of  $\approx 20 \div 100$ with respect to the adopted value or by adding a second electron component.
Both these possibilities seems quite artificial and thus an hadronic origin of the TeV emission seems favored based on spectral information only \cite{tanaka,morlino}. 

Recently, a number of order of magnitude estimates have been presented against hadronic models for the origin of the TeV emission from SNRs, and thus indirectly supporting the leptonic scenatio \cite{katz,butt,plaga}.
The main criticisms against the hadronic models can be summarized as follows, taking as reference case the SNR RX J1713.7-3946: {\it i)} in order to fit the TeV data within an hadronic model, the gas density cannot be too low (see Eq.~1). The high density would then imply a strong X-ray thermal Bremsstrahlung emission which is not observed \cite{katz}. {\it ii)} The high value of the magnetic field expected in hadronic models implies that a meager number of electrons is needed to explain the observed synchrotron X-ray emission. This translates into very low values of the ratio between CR electrons and protons $K_{ep}$ in the SNR, significantly smaller that the value $K_{ep} \sim 10^{-2}$ observed in the Galaxy \cite{katz,butt}. {\it iii)} There is no clear tight spatial correlation down to small angular scales between the gamma ray flux and dense molecular clouds, while this should be observed if gamma rays had a hadronic origin \cite{plaga}.

The issue of the missing thermal emission (problem {\it i)}) can be solved by reminding that the shock heating of the plasma is relevant only for ions, which carry the inertia of the flow.
Electrons can be heated downstream of the shock via Coulomb collisions with hot ions, but the characteristic time scale of the process is much too long to establish electron-ions temperature equilibrium \cite{spitzer,rakowski}.
Other processes, possibly mediated by plasma waves, might heat electrons and facilitate the equipartition (see e.g. \cite{rakowskihybrid}), but since the nature of such processes is uncertain, it is not possible to draw any firm conclusion.
It has to be noted that the plasma temperature can be further reduced if the shock is converting a large fraction of the available kinetic energy into CRs, since this has to happen at the expenses of a reduced gas heating \cite{donheating,abg}.
This would substantially reduce the thermal X-ray emission, since the peak of the emission would be shifted towards the UV energy range \cite{dorfi}.
For RX J1713.7-3946, if an electron temperature equal to one hundredth of the proton temperature is assumed, no thermal X-rays are expected \cite{morlino}. 

The low value of the electrons to protons ratio $K_{ep}$ (problem {\it ii)}) is not necessarily a problem for hadronic models.
In fact, the $K_{ep}$ value of a SNR at a given age does not need to agree with the value measured for local CRs.  
Low energy electrons are likely to be accelerated during the late phase of the SNR evolution, when the $K_{ep}$ value can be different from the present one \cite{tanaka}.
Also the lack of a good spatial correlation between gamma ray emission and the position of dense clouds (problem {\it iii)}) can be explained by remembering that molecular clouds are mainly neutral. This might reduce the level of magnetic turbulence (via damping) needed to have effective particle acceleration \cite{morlino,malkovcloud} and would worsen the spatial correlation between gamma rays and gas density.
Independently of the actual nature of the emission, the improving quality of multiwavelength data and the intrinsic non-linearity of the problem of efficient CR acceleration \cite{malkov,jones} make the above mentioned simplistic calculations inadequate and demand for a careful and detailed modeling \cite{donandme}.

At GeV energies, a number of spatial associations between EGRET sources and SNRs have been proposed \cite{sturner,esposito}, but despite extensive investigations, not a single SNR has been unambiguously identified by EGRET. However, a correlation was claimed to exist between unidentified EGRET sources and SNRs \cite{diegoreview}.
The detection of the SNR IC443, first proposed in Ref.~\cite{sturner} and recently confirmed by  the AGILE team \cite{tavani} constitutes the first clear evidence for a GeV SNR.
The detection of SNRs at GeV energies is of crucial importance for constraining emission models. 
An extended coverage of the gamma ray measurements from GeV to TeV energies will facilitate the distinction between hadronic and leptonic models, since they predict quite different spectral features (see Figs.~1 \& 2). In particular,  the detection of the slight spectral hardening of the spectrum expected in hadronic models \cite{volkRXJ,donandme}, would allow to test non linear shock models for CR acceleration, which predict concave spectra for the accelerated particles \cite{jones,malkov,pasquale,kang}.
Simulations of FERMI performances indicates that deep observations ($\sim 5$ yr) are needed to firmly detect TeV bright remnants such as RX J1713.7-3946 if the emission is hadronic \cite{funkSNRs}.

Finally, the detection of neutrinos would constitute an unambiguous signature for the presence of CR protons in SNRs. 
However, at least for RX J1713.7-3946, the neutrino spectrum is expected to cut off at an energy of few TeV, if the gamma ray emission is entirely due to hadronic interactions (Fig.~1, right panel). Unfortunately this is the energy region where neutrino telescopes reach their best performances.
Thus, a detection of SNRs in neutrinos, though possible, seems challenging \cite{kappes}.

\section{MOLECULAR CLOUDS: PROBES OF THE CR SPECTRUM THROUGHOUT THE GALAXY}

The importance of the detection of molecular clouds in gamma rays is widely recognized, especially in relation to the problem of the origin of CRs. Molecular clouds located in the vicinity of CR accelerators, such as for example SNRs, provide a dense target for CR hadronic interactions, amplifying the resulting gamma ray emission from neutral pion decay and making easier the identification of CR sources \cite{montmerle,casse}.
On the other hand, even in the absence of a nearby CR accelerator, molecular clouds are still expected to emit gamma rays at a certain level, due to the interaction of background CR that penetrate the cloud \cite{issa,felixclouds}.
According to both these scenarios, molecular clouds are {\it passive} gamma ray emitters, in the sense that they provide a dense target for interactions of CRs which are accelerated somewhere else \cite{felixclouds2}. 
It has also been suggested that particle acceleration can operate inside molecular clouds, due to the presence of strong magnetic turbulence that might effectively scatter particles \cite{dogiel}.

The hadronic gamma ray emission from a SNR might be enhanced if the supernova shock is overtaking a massive molecular cloud \cite{adv}. This is because the molecular cloud provides a thick target for proton-proton interactions and the related gamma ray emissivity is expected to scale linearly with the gas density (see Eq.~1).
For this reason, the observed  spatial associations between a few TeV bright SNRs and massive molecular clouds \cite{IC443,W28,HESSJ1745,CTB} is considered as a fact supporting the hadronic origin of the emission and thus also the idea that SNRs might be the sources of galactic CRs.
However, the connection between high gas density and enhanced gamma ray emissivity is not as straightforward as it might appear.
In fact, the gas constituting molecular clouds is mostly neutral, which means that the magnetic turbulence on which particles acceleration relies, can be effectively damped \cite{luke,brian}, potentially reducing both the acceleration efficiency and the maximum energy of accelerated particles.
This might have an important effect in suppressing the TeV gamma ray emission (see e.g. \cite{malkovcloud}). 
Thus, a tight spatial correlation between gas density and gamma rays is not necessarily to be expected, even if the TeV emission from SNRs has a hadronic origin.
In the future, high resolution gamma ray observations of molecular clouds overtaken by SNR shocks will be useful to assess the relative importance of these two crucial effects: the enhancement of the gamma ray emission due to high density, and its reduction due to damping of magnetic turbulence and subsequent suppression of the acceleration efficiency.

The association between SNRs and molecular clouds can also be used to determine the distance of the remnant itself, which is in most cases quite uncertain.
Conversely, the distance of a molecular cloud can be inferred from the measured doppler shift of CO lines, coupled with the knowledge of the galactic rotation curve \cite{blitz}. 
Notably, for the remnant RX J1713.7-3946, the possible correlation of TeV gamma rays with CO data obtained with the {\it NANTEN} sub-mm telescope \cite{fukui}, allows to estimate the distance of the object of about 1 kpc. Correspondingly, an age of $1000 \div 3000$ years can be deduced, which supports the idea that RX J1713.7-3946 is the remnant of the supernova that, according to Chinese historical records, exploded in 393 AD \cite{china}.

\begin{figure*}[ht]
   \centering
   \includegraphics[width=.7\textwidth]{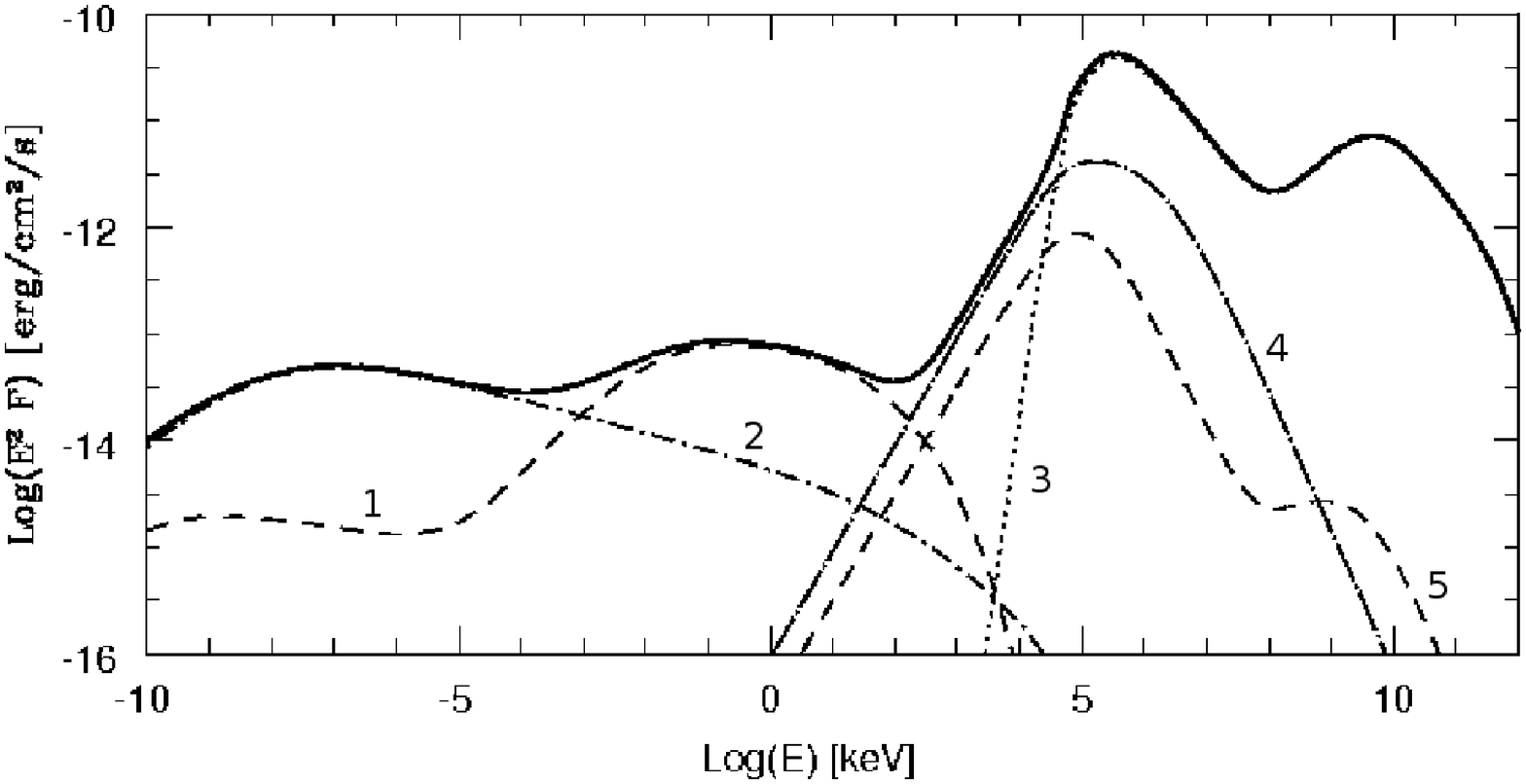}
   \caption{Broad band spectrum for a MC of mass $10^5 M_{\odot}$, radius $20$~pc, density $\sim 120$~cm$^{-3}$, magnetic field $20~\mu$G. The MC is at 100 pc from a SNR that exploded 2000 yr ago. Distance is 1 kpc. The dotted line shows the emission from $\pi^0$--decay (curve 3), from both background CRs and CRs escaping from the remnant. The dot-dashed lines represent the synchrotron (curve 2) and Bremsstrahlung (curve 4) emission from background CR electrons that penetrate the MC and the dashed lines the synchrotron (curve 1) and Bremsstrahlung (curve 5) emission from secondary electrons, respectively. Figure from Ref.~\cite{meHD}.}
\end{figure*}

Even if the molecular cloud is not overtaken by a SNR shock but it is located at a given distance $l_{cl}$ from the remnant, it can still be illuminated by CRs that escape from the SNR and produce there gamma rays \cite{atoyan,meapj}. 
For this scenario, it has been shown that, for typical SNR parameters and for a distance $d = 1$~kpc, a molecular cloud of mass $10^4 M_{\odot}$ emits TeV gamma rays at a detectable level if it is located within few hundred parsecs from the SNR \cite{meapj}. In this case the angular displacement between the remnant and the gamma ray emission is $\approx 6^{\circ} (d/1~{\rm kpc})^{-1} (l_{cl}/100~{\rm pc})$. 
This translates into the fact that sometimes the association between a gamma ray bright molecular cloud and the SNR which is accelerating the CRs responsible for the emission can be not so obvious, given  that the separation between the two objects can be bigger than the instrument field of view. 
More massive clouds ($\approx~10^5 M_{\odot}$ or above) irradiated by CRs might be detected up to the distance of the Galactic centre or beyond.

Non thermal emission from radio to X-rays is also expected from molecular clouds irradiated by CRs coming from a nearby SNR, due to synchrotron and Bremsstrahlung radiation from secondary electrons produced during the same hadronic interactions responsible for the production of gamma ray photons \cite{meHD,protheroe}. 
In Fig.~3 (from Ref.~\cite{meHD}), as an illustrative example, the broad band emission from a molecular cloud is plotted from radio waves to gamma rays. The cloud has a mass of $10^5 M_{\odot}$, a magnetic field of 20 $\mu$G and is located 100 pc away from a SNR which is converting 30\% of its total mechanical energy ($\sim~10^{51}$ erg) into CRs. The age of the SNR is 2000 yr and the distance to the observer is 1 kpc.
The contributions to the total radiation from background CRs, CRs escaped from the nearby SNR and secondary products of CR hadronic interactions in the cloud are plotted, as indicated in the figure caption. 
It is clear that the gamma ray emission by far exceeds the emission in the other energy bands. 
Following this rationale, it was proposed that some of the unidentified TeV sources detected so far, with no obvious or very weak counterparts in other wavelenghts (the so called "dark sources") \cite{HESSscan,HESSscan2,MILAGROsources}, might be in fact associated with massive molecular clouds illuminated by CRs \cite{meapj}.
Such a scenario is supported by the fact that many of the unidentified TeV sources are spatially extended, as molecular clouds are expected to appear.
The same interpretation has also  been suggested for the TeV gamma ray emission detected from a molecular cloud associated with the SNR IC 443 \cite{diegoIC443}.

In general, the hadronic gamma ray emission from a cloud illuminated by a nearby SNR accelerating CRs is expected to be the sum of two contributions: the gamma rays form interactions of CRs coming from the nearby SNR and the gamma rays from interactions of background CRs.
The former component is expected to exhibit a hard spectrum, reflecting the fact that low energy particles remain confined in the remnant for long time and thus cannot reach the cloud, while the latter will have a steep spectrum, reflecting the underlying proton spectrum of the background CRs  with slope $\sim E^{-2.7}$.
Thus, the gamma ray emission from a molecular cloud located in the proximity of a SNR might be characterized, at least at some stage of the remnant evolution, by a very peculiar spectrum which is steep at low (GeV) energies and hard at high (TeV) energies \cite{meHD}. 
This can be clearly seen in Fig.~3, where the low energy bump in the gamma ray emission, peaking at $\approx$ 100 MeV - 1 GeV energies, is the result of the interaction of background CRs, while the high energy bump, extending up to TeV energies and beyond, is due to interactions of CRs coming from the SNR.
The possibility of detecting sources with such a spectrum is also relevant for the issue of identifying GeV and TeV unidentified sources.  
One of the criteria generally adopted to support an association between a GeV and TeV source is, beside the positional coincidence, the spectral compatibility \cite{funkolaf}.
Such a criterium would not be applicable to the case described here.

Let us now consider the case of an isolated molecular cloud, with no CR sources located in its proximity.
In this case the molecular cloud is embedded in the ``sea'' of Galactic CRs, assumed to be constant throughout the Galaxy. If CRs can freely penetrate into the cloud, then the total gamma ray flux above a given energy $E_{\gamma}$ is solely determined by the mass of the cloud and its distance to the Earth and reads \cite{felixclouds}:
\begin{equation}
F_{cl}(\ge E_{\gamma}) \sim 1.45 \times 10^{-13} E_{TeV}^{-1.75} \left( \frac{M_5}{d_{kpc}^{2}} \right) {\rm cm^{-2} s^{-1}} 
\end{equation}
where $M_5$ is the cloud mass in units of $10^5 M_{\odot}$ and $d_{kpc}$ is the cloud distance in kpc.
This means that, since the mass and the distance of a cloud can be determined by means of observations of CO lines (see e.g. \cite{blitz,dame}), gamma ray observations of molecular clouds can be used to probe variations of the CR spectrum in different regions of the Galaxy.
The condition for the detectability of passive clouds embedded in the CR background with EGRET is $\delta ~ M_5/d_{kpc}^2 > 10$, where $\delta$ represents an enhancement (or suppression) of the CR density with respect to the local value. The EGRET diffuse gamma ray emission from the direction of prominent and nearby molecular clouds has been analyzed,  leading to a rough agreement with the expactations from models that assume a constant CR spectrum throughout the Galaxy ($\delta \approx 1$) \cite{digel}.
However, since there are only a few nearby clouds in the Galaxy satisfying this condition, a more sensitive instruments like FERMI is needed for these studies.

At TeV energies, several molecular clouds have been detected in the galactic centre region \cite{hessridge} or in the neighbourhood of a few SNRs \cite{IC443,W28,HESSJ1745,CTB}.
In the cases in which the mass and the distance of the cloud could be measured, it has been possible to evaluate the spectrum and the normalization of the CRs at the position of the cloud, by assuming that the emission has an hadronic origin.
Such an assumption is very well justified, especially in the case of the galactic centre ridge, because a good spatial correlation is observed between TeV emission and gas density, and because the few-degree-scale extension of the emission is hard to be explained as emission from electrons, since thay suffer severe synchrotron losses and thus cannot propagate large distances away from the acceleration regions. 
For the galactic centre region an enhancement in the CR density of a factor of $\delta = 3 \div 9$ with respect to the local CR spectrum has been inferred from gamma ray observations \cite{hessridge}, while for the molecular clouds detected in the field of the SNR W28 the inferred overdensities of CRs lies in the range $\delta = 10 \div 30$ \cite{W28}. This overdensities refer to CRs with energy above 1 TeV, which are the ones producing the observed gamma ray emission above 100 GeV.
In both cases, the inferred spectrum of the CRs is harder than the one observed locally.
These results demonstrate that there are places in the Galaxy, as the galactic centre ridge, where the CR spectrum can significantly differ from the local one.
 
More generally, a criterium for the detectability of molecular clouds with Cherenkov telescopes can be derived by recalling that such objects in general have an angular extension significantly larger than the typical instrumental angular resolution which is of the order of $\theta_{PSF} \approx 0.1^{\circ}$. 
For a Cherenkov telescope like H.E.S.S. the sensitivity at 1 TeV for an extended sources of angular size $\theta_s$ after 50 hours of exposure is roughly $\approx 10^{-13} (\theta_s/\theta_{PSF})$ TeV/cm$^2$/s. 
A cloud with mass $10^5 ~ M_5 ~ M_{\odot}$ and typical (and, for simplicity here, uniform) density $\sim 100$ cm$^{-3}$ located at a distance $d_{kpc}$ kpc will appear as a very extended gamma ray source with angular size $\theta_s \approx 1^{\circ} M_5^{1/3}/d_{kpc}$.
By combining this information with Eq.~2, and recalling that the quantity $\delta$ represents the excess of multi-TeV CRs with respect to the local value, it is possible to obtain a condition for the detectability that reads:
\begin{equation}
d_{kpc} < ~ 0.1 ~ \delta ~ M_5^{2/3} ~ \left( \frac{\phi^{TeV}}{\phi^{TeV}_{HESS}} \right)^{-1} \left( \frac{\theta_{PSF}}{0.1^{\circ}} \right) 
\end{equation}
if expressed as a function of the point source sensitivity $\phi^{TeV}$ of a generic instrument in units of the point source sensitivity of H.E.S.S. $\phi^{TeV}_{HESS}$.
This is a very rough order of magnitude estimate, but it can still be used as a rule of the thumb to assess the capabilities of Cherenkov telescopes in detecting extended molecular clouds.

It is clear from Eq.~3 that the detection of TeV gamma rays from massive ($M_5 \approx 1$) molecular clouds embedded in the galactic CR background ($\delta \approx 1$) is beyond the capabilities of currently operational Cherenkov telescopes like H.E.S.S.. 
This is in line with the fact that the molecular clouds detected so far in TeV gamma rays are very massive ($M_5 >> 1$) and/or located in regions characterised by an excess in the CR density ($\delta >> 1$).
Next generation instruments, such as CTA \cite{cta} and AGIS \cite{agis} , with an expected sensitivity improved by about one order of magnitude will detect more distant and weaker molecular clouds, and possibly nearby clouds even in the less optimistic case when $\delta = 1$.
This will lead to an improved knowledge of the distribution of CRs in the Galaxy and will possibly provide stringent constraints for CR propagation models.

Of course, what is discussed above is valid only under the assumption that CRs freely penetrate the clouds. 
The issue of the penetration or exclusion of CRs from clouds has been investigated in several papers \cite{cesarsky,skilling,dogiel}, in which quite different conclusions have been drawn, going from the almost free penetration to the exclusion of CRs up to tens of GeV.
The exclusion of CRs from clouds can suppress the expected gamma ray emission and lead to a spectral hardening towards the centre of the cloud, due to the fact that low energy CRs are excluded more effectively than high energy ones \cite{mebarca}.

\section{THE DIFFUSE GAMMA RAY EMISSION \\ FROM THE GALACTIC PLANE}

The diffuse GeV emission detected by EGRET along the galactic plane is believed to be mainly due to the decay of neutral pions produced in hadronic interactions between CRs and interstellar matter.
A strong point in favor of this hypothesis is the fact that the diffuse GeV emission correlates with the spatial distribution of the interstellar gas which constitutes the target for CR interactions \cite{EGRETdiffuse}.
Below $\sim 100$ MeV inverse Compton and electron Bremsstrahlung become the main contributors to the gamma ray emission.

Several attempts to model the diffuse galactic gamma ray emission have been developed so far, from simple earlier models where the CR flux was assumed to be constant throughout the Galaxy \cite{berezinsky} or determined by using dynamical balance arguments \cite{bertsch}, to numerical studies of CR propagation in the Galaxy, tuned in order to match the CR spectrum observed locally \cite{galprop}. 
Remarkably, all these models reproduce quite well both the level and the spectrum of the diffuse emission with the notable exception of the so called "GeV excess", where data exceed model predictions by roughly a factor of 2. 
The GeV excess can be explained by relaxing the assumption that the spectrum of CRs observed locally (nuclei, electrons or both) is representative for the whole Galaxy.
Data can be indeed fitted by assuming that the actual CR proton spectrum in the Galaxy is harder than the one measured at Earth \cite{wolfendale,aa2000}.
Another possibility is to assume a hard spectrum of CR electrons or that the typical CR electron intensity in the Galaxy is significantly higher than that measured locally \cite{galprop}. This would make the inverse Compton contribution to the total diffuse gamma ray emission dominate over the hadronic one at photon energies above $10 \div 100$ GeV.

Recently, the MILAGRO collaboration reported on the detection of multi-TeV diffuse gamma-ray emission from two regions in the Galactic plane: the Cygnus region (located at galactic longitude $65^{\circ} < l < 85^{\circ}$) \cite{milagrocyg} and the portion of the inner Galaxy visible from the location of MILAGRO ($30^{\circ} < l < 65^{\circ}$) \cite{milagroinner}.
When placed alongside the above mentioned
lower energy measurements by EGRET, such detections provide an interesting insight  
into both the spatial distributions of CRs
in the Galaxy and their spectrum over a broad energy range. 
Puzzlingly, conventional CR propagation models that reproduce quite fairly at least the observed level of the GeV emission (with the exception of the GeV-excess), fail to reproduce the MILAGRO observations by about one order of magnitude in the case of the Cygnus region \cite{milagrocyg} and a factor of $\sim 5$ for the inner Galaxy \cite{milagroinner}.
However, it has to be kept in mind that such models are designed to describe the global gamma-ray emission from the Galaxy, and are not expected to reproduce correctly local excesses in the gamma ray emission due, for example, to the presence of localized CR sources. 
Thus, the discrepancy between models predictions and MILAGRO data is representative of the fact that the CR spectrum measured locally is probably not representative for the whole Galaxy.
As done in the case of the GeV excess, a harder spectrum of CR protons or electrons can be invoked to solve the problem.

Thus, the origin of the multi-TeV emission detected by MILAGRO from the Cygnus region and from the central Galactic plane region remains unclear.
Either leptonic or hadronic processes have been proposed to be the dominant contributors to the detected gamma-ray flux, 
with the possibility of a transition from a dominant hadronic contribution to a leptonic one in the GeV-TeV 
energy range \cite{galprop}.
In the case of the hadronic scenario, one would expect, as is observed, a narrow extension in latitude of both the EGRET \cite{EGRETdiffuse} and MILAGRO emission \cite{milagroinner}. In this case the extension of both GeV and multi-TeV emission is determined only by the gas distribution in the Galactic disk, which constitutes the target for proton--proton interactions.
In the leptonic scenario, the multi-TeV emission is produced via inverse Compton scattering by $\approx 100$~TeV electrons which can propagate over a distance of $\approx 100$~pc before being cooled by synchrotron and inverse Compton losses \cite{aa2000}. Thus, also in this case a narrow latitude distribution of gamma rays is expected, if the sources of cosmic ray electrons are concentrated around the galactic plane.
In this case, however, the extension of the gamma-ray emission would be energy dependent, since higher energy electrons are cooled faster and can propagate shorter distances \cite{aa2000}.  
Thus, an accurate comparison of the extension in latitude of the emission detected at GeV and multi-TeV energies is of crucial importance, and future observations with improved angular resolution by FERMI and HAWK might help in discriminating between the two scenarios. 

The fact that both the Cygnus region and the inner Galaxy are characterised by an enhancement in the gas density might suggest an hadronic origin for the diffuse multi-TeV emission detected by MILAGRO.
Moreover, in the case of the Cygnus region, the morphology of the multi-TeV emission correlates with the CO emission, which traces the gas density \cite{milagrocyg}.
Similarly, for the region close to the inner Galaxy, the rather narrow latitude profile of the multi TeV-emission, concentrated around the dense Galactic disk, might suggest a hadronic interpretation of the gamma-ray data, though also a leptonic one seems feasible \cite{milagroinner}.
Since the production of gamma rays is accompanied by a corresponding neutrino flux only in the hadronic scenario, 
the search for a diffuse flux of neutrinos from the inner Galaxy, which might be detectable by km$^3$-size neutrino detectors, will provide a unique opportunity to disentangle the origin of the multi-TeV radiation \cite{mediffuse}.

\section{OBSERVATIONS IN THE MULTI-TeV RANGE: ARE SNRS COSMIC RAY PEVATRONS?}

The observed CR spectrum extends as a featureless power law 
up to the {\it knee} at an energy of a few PeV, where it undergoes a significant steepening.  
This suggests that the sources of galactic CRs, whichever they are, must be able to accelerate particles up to {\it at least} a few PeV. 
As seen in Sec.~2, both the X-ray observations of SNRs \cite{bamba,vink,yasunobu} and the recent theoretical studies of magnetic field amplifications due to CR streaming at shocks \cite{lucek,bell} seem to suggest the presence of a strong magnetic field with intensity $\approx 100 ~ \mu$G or more, which is the value needed to reach (or even exceed) PeV particle energies in diffusive shock acceleration.

A decisive and unambiguous indication for the acceleration of PeV protons 
in SNRs can be provided by observations of $\gamma$-rays at energies 
up to 100 TeV and beyond. Because of the Klein-Nishina effect 
the efficiency of inverse Compton scattering 
in this energy band is dramatically reduced. Thus, the possible detection of SNRs in gamma rays up to these energies could be unambiguously interpreted as an evidence for the fact that such objects act as CR PeVatrons \cite{meapj}. 
Although the potential of the current ground-based 
instruments for detection of such energetic 
$\gamma$-rays is limited, it is expected that 
the next generation arrays of imaging Cherenkov telescopes 
optimized in the multi--TeV energy range will become 
powerful tools for this kind of studies \cite{tenten,felixfuture}. 

It should be noted that the number of SNRs currently bright in $> 10$ TeV $\gamma$-rays 
is expected to be rather limited. Multi--PeV protons can be 
accelerated only during a relatively short period of the SNR evolution, namely, at the end of the free--expansion phase/beginning of the Sedov phase, when the shock velocity is high enough to allow sufficiently high acceleration rate. When the SNR enters the Sedov phase, the shock gradually slows down and correspondingly the maximum energy of the particles that can be confined within the remnant decreases. This determines the escape of the most energetic particles from the SNR \cite{ptuskin}. Thus, unless our theoretical understanding of particle acceleration at SNRs is completely wrong, we should expect an energy  spectrum of CR inside the SNR approaching  
PeV energies only at the beginning of the Sedov phase, typically 
for a time $< 1000$ years. 
When the SNR enters the Sedov phase, the high energy cutoffs in the spectra of 
both protons and $\gamma$-rays gradually moves to lower energies, while the highest energy particles 
leave the remnant \cite{ptuskin}. 
This can naturally explain why the $\gamma$-ray spectrum of the best studied SNR 
RX J1713.7--3946 above 10 TeV becomes rather steep with photon index $\approx 3$
\cite{RXJ100TeV}.

The detection of multi--TeV $\gamma$-rays 
generated by the CRs that escape the SNR might constitute the smoking gun for particle acceleration up to PeV energies \cite{meapj}. 
A molecular cloud located close to the SNR can provide an effective 
target for production of $\gamma$-rays. 
The highest energy particles ($\sim$ few PeV) escape the shell 
first. Moreover, generally they diffuse in the interstellar medium 
faster than low energy particles. Therefore they arrive first to the cloud,
producing there $\gamma$-rays (and also neutrinos) with very hard energy spectra.  
Note that an association of SNRs with clouds is naturally 
expected, especially in star forming regions \cite{montmerle}. 
The duration of $\gamma$-ray emission in this case is determined by the time 
of propagation of CRs from the SNR to the cloud.
Therefore $\gamma$-ray emission of the cloud lasts much longer
than the emission of the SNR itself. This makes the detection of 
delayed $\gamma$-ray and neutrino signals more probable.
The detection of these multi--TeV $\gamma$-rays from 
nearby clouds would thus indicate that the nearby SNR 
in the past was acting as an effective PeVatron \cite{meapj}.
Both $\gamma$-rays and neutrinos are emitted with fluxes detectable by currently operating and forthcoming instruments.  
Since the $\gamma$-ray spectra from clouds are extremely  hard, $\gamma$-ray telescopes operating at very high energies ($> 10$ TeV) would be the best instruments for this kind of study. 
Remarkably, a detection of such emission would not only reveal the acceleration of PeV CRs, but also suggest the best targets for neutrino observations, since neutrino telescopes are optimized for observations in the multi--TeV energy region.

\section{CONCLUSIONS AND FUTURE PERSPECTIVES}

The recent developments of gamma ray astronomy, especially at very high (TeV) energies, constitute an important step towards the solution of the problem of the CR origin.
The detection of SNRs in TeV gamma rays by Cherenkov telescopes allows us to study several aspects of particle acceleration in these objects with unprecedented accuracy (see e.g. \cite{rev:felix}). 
On the other hand, the observation of diffuse TeV emission from the galactic plane by air shower detectors such as MILAGRO \cite{milagroinner}, when combined with earlier EGRET observations of the diffuse GeV emission \cite{EGRETdiffuse}, provide useful information on the spatial distribution of CRs in the Galaxy and on their spectrum.
The combination of pointed observations performed with Cherenkov telescopes and the continuous monitoring of the whole sky by air shower detectors revealed to be a very successfull approach, and the complementarity of these two techniques will hopefully be exploited in the future by instrument like CTA \cite{cta}, AGIS \cite{agis} and HAWC \cite{hawc}.

Though both morphological and spectral studies of SNRs seem to favor an hadronic origin of the gamma ray emission, leptonic models still cannot be considered ruled out.
The detection of neutrinos from the direction of SNRs would unambiguously solve the issue, and prove that SNRs can indeed accelerate CR protons \cite{dav,vissani,kappes,halzen,meapj}.
However, since such detections appear to be challenging even for km$^3$-scale neutrino telescopes, the search for evidence of CR proton acceleration coming form gamma ray observations is mandatory.
Besides the issue of the hadronic or leptonic origin of the gamma ray emission from SNRs, the problem of the (unexpected) high level of isotropy observed from CRs up to the energy of the knee still persists (e.g. \cite{hillas}). 

The solution of the long standing problem of the origin of galactic CRs might finally come from an improved knowledge of the CR spectrum and spatial distribution throughout the Galaxy and from a better understanding of several aspects of the problem, including acceleration of CRs at shocks, escape of CRs from SNRs and their propagation in the Galaxy.

\subsection{Future perspectives: theory}

The non linear theory of diffusive shock acceleration gives us a satisfactory description of how particles are accelerated at shocks and has been successfully applied to model the multiwavelength emission from SNRs (see e.g. \cite{volkRXJ,donandme}).
Despite the success of the theory, some issues still need further investigation.
In particular, new aspects of the basic physics of the CR-driven magnetic field amplification at shocks have been understood \cite{bell}, but a fully self consistent treatment of the problem is still missing.
A better understanding of this mechanism might provide a solid theoretical interpretation for the very high values of the magnetic field inferred from X-ray observations of SNRs \cite{bamba,yasfirst,vink,yasunobu} and will allow us to estimate the particle acceleration rate and thus the maximum particle energy achievable within this theoretical framework \cite{bell}.

Another missing piece of information is the way in which accelerated CRs leave the SNR.
While protons are believed to be released gradually during the whole Sedov phase as the SNR shock slows down \cite{ptuskin}, electrons are trapped within the remnant due to severe synchrotron losses and are probably released in the interstellar medium solely during the late phase of the SNR evolution.
Surprisingly, this aspect of the problem has not yet received as much theoretical attention as it deserves. Studies in this direction will allow us to estimate the spectrum of CR protons and electrons injected in the interstellar medium and will possibly tell us something about how the proton to electron ratio varies during the SNR evolution. 

After being released into the interstellar medium, CRs propagate in the galactic magnetic field.
The propagation of CRs is generally investigated by means of numerical codes that solve the transport equation given the diffusion coefficient and possibly the structure of the galactic magnetic field.
The most popular amongst these codes is GALPROP \cite{GALPROPreview}, which provides as an output also the diffuse gamma ray emission from CR interactions in the interstellar gas and radiation field \cite{galprop}.
The main limitation of such approaches is that they normally assume a continuous and spatially smooth injection of particles along the galactic plane and, being normalized to local CR data, cannot account for localized excesses of CR sources.
These limitations became evident when attempts to use GALPROP to model the diffuse emission detected by MILAGRO from the Cygnus region failed  to fit observations by about an order of magnitude \cite{milagrocyg}.
Thus, though GALPROP-like approaches can predict global properties of the diffuse gamma ray emission from the Galaxy, they cannot be applied to model local excesses of the diffuse gamma ray emission.
 
One possible way to improve the studies of CR propagation and of the related diffuse emission is to include in numerical codes the stochasticity of SNR explosions, which would lead to the appearance of localized (both in time and space) excesses in the CR density close to their sources \cite{atoyan,meapj,busching}. The expected association between supernovae and dense gas regions that provide a thick target for CR hadronic interactions might further enhance the resulting gamma ray emission.
Obviously, this approach, being statistical in nature,  would only allow to infer average properties of the gamma ray emission from the Galaxy, such as, for example, the number of molecular clouds irradiated by CRs detectable in gamma rays as unidentified sources \cite{meHD}, or the amplitude and the spatial scale of the fluctuations in the diffuse gamma ray emission.
These predictions would provide clear observational tests for propagation models and for the SNR hypothesis for CR origin.

Finally, self consistent propagation models where CRs propagate away from their sources due to scattering on the magnetic turbulence they themselves produce \cite{vladvlad} need to be further developed and, together with the knowledge of the CR spectrum injected by SNRs in the interstellar medium, might clarify the issue of the isotropy observed in the arrival direction of CRs up to PeV energies. 

\subsection{Future perspectives: observations}

Future observations in the gamma ray domain are expected to drive the progresses in the field.
Comprehensive reviews of the expected performances and scientific outputs from next generation gamma ray instruments can be found in Refs.~\cite{felixfuture,whitepaper,rev:felix}.
Here the expected contribution from these observations to the problem of CR origin is summarized.

First of all, it has to be noted that observations in different energy regions will provide insights into different aspect of the problem. Following previous classifications, we divide the gamma ray spectrum into three regions: {\it i)} the GeV domain, which spans the energy spectrum in the interval $\sim$ 100 MeV -- 100 GeV, currently covered by the FERMI satellite; {\it ii)} the TeV domain, where Cherenkov telescopes and air shower detectors operate, $\approx$ 100 GeV -- 10 TeV, and {\it iii)} the multi-TeV domain, approximatively $>$ 10 TeV, which will be explored by next generation of instruments.

\subsubsection{GeV Energy Domain: FERMI}

FERMI is now scanning the whole sky at GeV energies with unprecedented sensitivity.
The recent detection of the SNR IC443 by AGILE suggests that FERMI, with its better sensitivity will possibly detect several of these objects.
A detection of SNRs at GeV energies is of crucial importance to discriminate amongst hadronic and leptonic origin of the gamma ray emission.
Moreover, FERMI will measure the diffuse emission from the galactic plane, and thanks to its improved angular resolution is expected to reveal structures in the diffuse emission which reflect the underlying clumpiness of the interstellar medium.
The possible detection of individual molecular clouds will allow us to probe the CR spectrum in different parts of the Galaxy, providing stringent constraints to CR propagation models and revealing the presence of localized excesses of CR sources.

\subsubsection{TeV Energy Domain: how many sources?}

Next generation of instruments operating in the TeV domain are expected to increase significantly the number of sources detected along the galactic plane. 
The improved angular resolution of Cherenkov telescopes will allow better morphological studies of extended galactic sources, useful to search for correlations between gamma ray emission and gas density and/or X-ray emission. 
The presence or absence of such correlations will help in discriminating between hadronic and leptonic models for the gamma ray emission.
Air shower detectors will complement such studies by mapping the extended emission from the whole Galaxy. 

\subsubsection{Multi-TeV Energy Domain: searching for Cosmic Ray PeVatrons}

Most of the galactic sources detected so far by Cherenkov telescopes do not show evidence for a high energy cutoff in their spectra \cite{HESSkarl}.
For this reason, an extension of the observed energy range in the multi-TeV domain is desirable, since it will allow us to search for cutoffs in such spectra, thus probing the most extreme particle accelerators in the Galaxy.
Moreover, the sources of CRs with energies up to the knee are expected to exhibit a cutoff in this domain of the gamma ray spectrum. 
This implies that observations in the multi TeV region might finally reveal the nature of the CR PeVatrons.

\subsubsection*{Acknowledgments}
The author thanks the organizers of the 21$^{st}$ European Cosmic Ray Symposium, and F. Aharonian and A. Taylor for useful comments on the manuscript.
He also acknowledges support from the European Community under a Marie Curie Intra--European--Fellowship.



%

\end{document}